# High power, electronically-controlled, source of user-defined vortex and vector light beams based on a few-mode fibre amplifier

Di Lin,[1,*] Joel Carpenter,[2] Yutong Feng,[1] Yongmin Jung,[1] Shaif-ul Alam,[1] David J. Richardson[1]

[1]*Optoelectronics Research Centre, University of Southampton, SO17 1BJ, UK*
[2]*School of Information Technology and Electrical Engineering, The University of Queensland, Brisbane, Queensland 4072, Australia*
*\*Corresponding author: di.lin@soton.ac.uk*

**Abstract:** Orbital angular momentum (OAM) based structured light beams provide an additional degree of freedom for practical applications ranging from optical communication to laser-based material processing. Many techniques exist for generating such beams within laser sources and these primarily rely upon the use of specially designed optical components that limit laser power scaling and ready tunability of the topological charge and polarization of the output OAM beams. Here we show that some of these limitations can be overcome by employing a computer controlled reflective phase-only spatial light modulator (SLM) to adaptively tailor the input (and subsequent output) beam wavefront and polarization in a few-mode fibre amplifier. In this way modal-coupling induced beam distortion within the fibre amplifier can be mitigated and we are able to generate at will any desired supported spatial mode guided in the fibre, including conventional LP modes, scalar OAM modes and cylindrical vector modes, at average powers >10 W and with a peak power of >11 kW. Our results pave the way to the realization of practical high-power structured laser sources with tunable chirality and polarization.

**Introduction**

Structured laser beams have recently become a topic of significant interest due to the fact that they provide additional degrees of freedom in terms of optical phase, polarization and amplitude and have subsequently found many novel applications in, amongst others, the optical manipulation of particles [1, 2], optical free space and fibre based communications [3, 4], laser-based material processing [5, 6], particle acceleration [7] and microscopy [8, 9]. Foremost among the family of structured laser beams are scalar vortex beams and cylindrical vector beams. Scalar vortex beams are characterized by an annular intensity profile associated with an azimuthally varying phase structure of $\exp(il\varphi)$ (where $l$ is the integer of topological charge) so that they carry OAM. They are typically generated using dynamic phase approaches by imposing a helical phase front on an incoming Gaussian beam using a phase plate [10], computer generated hologram [11], or phase-only spatial light modulator (SLM) [12]. On the other hand, cylindrical vector beams are associated with $l$-dependent azimuthally varying polarization distributions (such as radial and azimuthal polarizations), which can be interpreted as combinations of OAM and spin states that result in a plane wavefront that does not carry OAM [13]. Such beams are created with geometric phase approaches by employing a q-plate [14], s-waveplate [15] or meta-surfaces [16] to transform an incident beam with homogenous polarization (e.g. linear polarization) into a beam with the desired vector polarization state. Whilst these passive beam conversion approaches are relatively easy to implement, they tend to suffer from relatively low modal purity, low conversion efficiency and/or limited power handling capability.

Over the past decade, remarkable advances have been made in the direct generation of such structured light beams from gain media including bulk solid-state lasers/amplifiers [17-19], fibre lasers/amplifiers [20, 21], organic lasers [22] and gas lasers [23], offering advantages of higher modal purity and much higher power scalability. However, such "structured light" laser systems can typically only generate a specific fixed spatial mode. The selective and flexible generation of such beams (i.e. tuning of the polarization and topological charge of the OAM states on demand), ideally from a compact laser source at high output powers with high modal purity and with simultaneous control of the temporal characteristics, is of great interest but remains a great challenge. A few methods that provide some flexibility in switching the laser output intensity profiles, or tuning the chirality of the OAM states, have been implemented in bulk solid-state lasers [24-26] and semiconductor based microlasers [27-30] but generally the performance has been limited by damage to the special optical components used due to the high intracavity powers involved. One promising way to overcome this limitation is to exploit the master-oscillator power amplifier (MOPA) approach that amplifies a reconfigurable structured light seed beam through an appropriate gain medium with a high gain whilst preserving a high mode purity. We have recently demonstrated a proof of concept demonstration that flexibly generates various structured light beams from a compact multicore fibre amplifier by coherently combining the individual Gaussian-shaped output beamlets with appropriate complex amplitude [31]. However, the unavoidable relatively high loss (typically ~40%-50%) associated with the non-colinear beam combination might ultimately hinder its use in the high power regime.

The vector nature of the diversity of electric field distributions of the eigenmodes guided in a cylindrically symmetric isotropic large core, few-mode optical fibre makes this an attractive compact platform for the dynamic generation of different structured spatial modes at high powers given the high efficiency and high power scalability of traditional Gaussian-shaped beams in such gain fibres. However, the vector modes with the same radial and azimuthal index are either strictly degenerate with exactly the same effective refractive indices (i.e. $\text{EH}_{l-1,m}^{\text{even}}$ and $\text{EH}_{l-1,m}^{\text{odd}}$), or nearly degenerate with a small difference in the effective refractive indices ($\Delta n \sim 10^{-5} - 10^{-7}$, i.e. $\text{EH}_{l-1,m}^{\text{even}}$ and $\text{HE}_{l+1,m}^{\text{even}}$). Thus even the slightest perturbations to the cylindrical symmetry of the fibre (i.e. bends, twists, and refractive index inhomogeneities) can cause strong coupling between such modes, resulting in the formation of the so called scalar LP modes that are usually observed at the output of fibres/fibre devices [32, 33]. Although, various mode excitation/selection techniques have been demonstrated in fibre lasers/amplifiers, these have generally only allowed the generation of a single, non-tailorable lowest order vortex or vector mode for a given laser cavity design [21, 34, 35]. In such systems, mode coupling induced distortion of the vortex or vector beam can be reduced, to some extent at least, by deliberate mechanical manipulation of the fibre e.g. by applying extra pressure, bends, or twists at certain points to ensure a doughnut-shaped intensity profile at the fibre output. These mitigation measures though are wholly empirical, typically lack repeatability and become increasingly ineffective at higher power levels and for fibres supporting a larger number of guided modes. This ultimately hinders the utilization of such exotic laser sources for end-users who are not laser experts. More advanced fibres with annular index profiles that reduce mode coupling and support the stable propagation and amplification of vortex and vector modes over long distance have been proposed and demonstrated [33, 36, 37]. However, such fibres rely on critical design and rigorous control of the refractive index profile and thickness of the annular core required and are characterized by a small effective mode area. Thus fibre fabrication is challenging and the tight mode confinement is not compatible with high power laser operation.

Here, we overcome the aforementioned limitations through the use of an adaptive input wavefront and polarization shaping technique which allows for precise control of the amplification of structured spatial modes in a commercially available few-mode large mode area (FM-LMA) fibre. A reflective phase-only SLM is employed to excite and correct the wavefront and polarization of the input beam to a final FM-LMA fiber amplifier stage in a picosecond pulsed MOPA system to obtain the desired structured spatial modes at the MOPA output. We experimentally demonstrate the computer-controlled generation of a variety of spatial modes supported by the FM-LMA fibre, obtaining very good modal purity (>90 %), an average power of >10 W and a corresponding peak power of 11 kW. This includes the arbitrary generation of conventional scalar LP modes, linearly polarized OAM modes and vector eigenmodes.

**Results**

**Concept**

Light propagation through passive fibres remains deterministic even in the case of strong modal coupling. Indeed, a number of techniques based on manipulation of the incident wavefront (including digital phase conjugation [38], transmission matrix [39] and adaptive wavefront shaping [40]) have been demonstrated with great success to selectively generate a desired electric field distribution at the output of a passive MMF by using a phase only SLM [41-43]. Recently the iterative optimisation based adaptive wavefront shaping technique has been adapted to shape the light in FM-LMA fibre amplifiers and MMF amplifiers in which light transmission becomes nonlinear due to the gain competition among the transverse modes induced by non-uniform gain saturation. For instance, by employing a photonic lantern as the beam shaping element, the amplification of selected scalar LP modes in a LMA fibre amplifier has been demonstrated [44, 45], and amplification in a MMF amplifier resulting in a tightly-confined, single-spot (or multiple spots) output beam has been obtained using a deformable mirror to tailor the incident beam wavefront [46]. However, the deformable mirror has a very limited number of actuators which means it cannot readily be used to generate more sophisticated structured beams with spatially varying phase and/or polarization distributions. To date, no technique has been demonstrated with the capability of allowing the user to generate a specific selected complex guided spatial mode in a fibre laser/amplifier at high output power.

Spatial beam shaping with SLMs is widely used as one of the most versatile techniques for generating structured light beams in free-space with high fidelity, however there are issues with this approach. Firstly, shaping efficiency can be an issue particularly due to diffraction loss when there is a limited intensity overlap between the incident beam and the target beam, secondly liquid crystals devices have a relatively low damage threshold (typically 2 W cm$^{-2}$) and this limits the power level of the beams that can be reflected from these devices. Here, we propose a "digital fibre amplifier" [24] to overcome these issues. A schematic of the system is illustrated in Fig. 1 (the operation of the system is described in more detail in the Method Section). One of the central ideas behind our approach is to place the beam shaping SLM prior to a FM-LMA fibre amplifier, where the incident power levels are modest, allowing a pre-shaped beam to be amplified to a target beam with much higher output powers than could be applied directly to the SLM if it were used to shape the beam at the amplifier output, whilst also avoiding any shaping losses at this critical point in the system. The SLM provides a reprogrammable means of converting a Gaussian beam into a beam with arbitrary spatially dependent polarization. In the forward direction, the SLM system converts a pre-amplified Gaussian-shaped picosecond pulses (with a pulse duration of ~150 ps and a wavelength of 1035nm at a repetition rate of 5.9 MHz), into the required field which must be excited at the input of the fibre to generate the desired output. Any beam distortion and depolarization induced by mode coupling during the amplification is pre-compensated by adaptively shaping the incident wavefront and polarization state through an iterative optimisation process. The shaped incident wavefront can be encoded as a superposition of a basis set of orthogonally linearly polarized modes generated by the SLM with an appropriate complex amplitude:

$$\mathbf{E}_{\text{in}} = \sum_{l=-N}^{l=N} \left( a_l |H,l\rangle + b_l |V,l\rangle \right) \quad (1)$$

where $|H,l\rangle = e^{il\varphi}|H\rangle$ and $|V,l\rangle = e^{il\varphi}|V\rangle$ represent the basis modes (horizontal polarization $|H\rangle$ and vertical polarization $|V\rangle$) with a helical wavefront $e^{il\varphi}$ (and with an intensity profile matched to those of the fibre eigen modes), and $a_l = a_{l0}e^{i\theta_{l1}}$ and $b_l = b_{l0}e^{i\theta_{l2}}$ represent the complex amplitude of each mode (where $a_{l0}$ and $b_{l0}$ represent the normalized amplitude of the horizontal and vertical components, respectively; and $\theta_{l1(2)}$ represents the relative phase of each mode). Any spatial mode can be expressed as an appropriate superposition of these basis modes. A modified iterative Fourier transform algorithm is used to calculate the phase masks required to generate the intensity and phase distributions of each basis mode [47, 48]. The transformation efficiency of each basis mode varies from ~30% (for the fundamental Gaussian mode) to ~54% (for the first higher order OAM mode ($|l|$=1)). In addition, there is a static insertion loss of ~2.5dB for the beam-shaper (measured from the input SMF to the output facet of the other SMF), resulting in a total loss of ~-7.7dB to ~-5.2dB for re-shaping the input beam to excite the desired beam in the following FMF. At the output, an analyser SLM (used in the reverse direction to the beam shaping SLM) can be used as a correlation filter to detect the portion of the field in the desired state. It is digitally configured with a transmission function $T(r) = \Psi_{\text{tar}}^*(r)$ (where $\Psi_{\text{tar}}^*(r)$ represents the complex conjugate of the target mode in terms of both amplitude and phase) so that the on-axis intensity of the correlation signal in the far field is proportional to the power of the output beam within that target mode [49]. The on-axis signal power can be measured and used as the merit function of a standard steepest gradient descent algorithm to provide iterative feedback to adaptively shape the input wavefront and polarization. This is achieved by adjusting the relative power and phase of each basis mode in Eq. (1) [50]. Therefore, the output beam can be electronically controlled simply by displaying different (complex) phase masks on the SLM. Our calculations show that the fibre can theoretically guide 20 spatial modes (accounting for the two polarizations) if kept straight. In practice, however, as the fibre was coiled with a diameter of ~25 cm, only 12 spatial modes can be effectively guided (up to the $LP_{02}$ mode) due to the high bend loss sensitivity of the higher order modes.

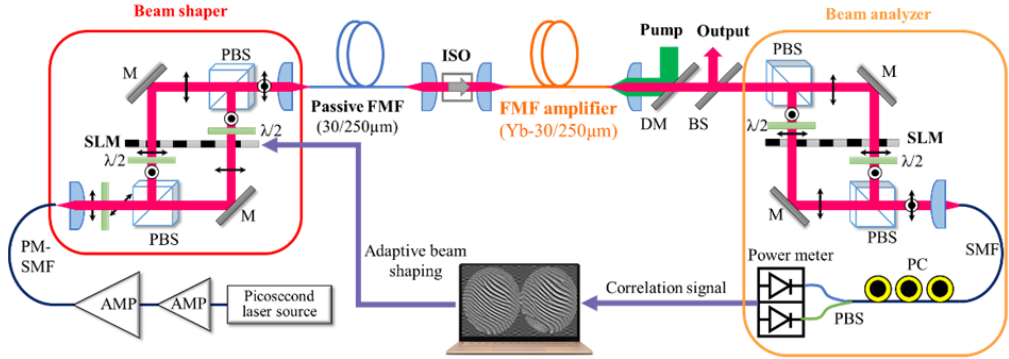

**Fig. 1 Digital fibre amplifier.** A linearly polarized Gaussian seed beam is tailored to excite an arbitrary spatial/polarization state at the input of the fibre amplifier by the incorporation of a SLM with some polarization diversity optics. The amplified output beam is characterized by a spatial mode correlation filter based on a second SLM that provides a feedback signal to the input SLM to enable adjustment of the excitation state until the target beam is obtained. AMP: amplifier; PM: polarization maintaining; SMF: single-mode fibre; PBS: polarization beam splitter; DM: dichroic mirror; BS: beam splitter; PC: polarization controller; M: mirror; λ/2: half wave plate; ISO: isolator.

**Generation of scalar LP modes**

In a preliminary experiment, we tested the amplification of several fibre eigenmodes without implementing any adaptive input wavefront shaping which imitated the scenario of amplification in a conventional few-mode fibre amplifier. In this case, the beam-generator selectively converted the input Gaussian beam into an individual selected single spatial eigenmode, which was then launched into the amplifier. **Fig. 2**(a) shows the measured intensity profiles of the amplified output beams and the corresponding input signal beams. We can see that the amplification of an input doughnut-shaped vector mode (i.e. $TM_{01}$ and $EH_{11e}$) always results in an output intensity profile similar to that of the corresponding LP mode due to the strong intra-modal coupling. Moreover, the amplification of the vertically polarized fundamental $LP_{01}$ mode and the $LP_{02}$ mode are also vulnerable to polarization degradation and beam distortion and hence suffer from a reduced modal/polarization purity in practice. The modal decomposition result for the $LP_{01}$ mode (as illustrated in **Fig. 2**(b)) shows that the modal purity degraded to ~73.8 % with a significant fraction of the power coupled to the orthogonal linear polarization, leading to a polarization extinction ratio (PER) of 8.2 dB at the amplifier output. The other component of power contained in higher-order modes is mainly attributed to imperfect input beam launching conditions – any small component of power coupled to the higher-order modes at the launch end will experience a relatively high gain due to the spatially dependent gain saturation which will predominantly effect the launched $LP_{01}$ mode.

We then tested the feasibility of our beam-shaping concept with a targeted vertically polarized fundamental $LP_{01}$ mode. In this case, only the section of the second SLM allocated to controlling the vertical polarization component of the beam was coded with a phase mask corresponding to the conjugated $LP_{01}$ mode such that the correlation signal can be detected by the SMF attached to the power-meter. This is then used as the merit function in the iterative optimisation process. The initial phase of each mode was preset to 0 and the amplitude of each mode was randomly preset but with a higher value for the target $LP_{01}$ mode. The preset parameters are not critical in determining whether we can find a good solution or not, but a good choice can reduce the optimization time which typically takes about ~10-15 minutes. **Fig. 2**(c) shows the measured intensity profile of the optimised $LP_{01}$ mode, and the modal decomposition result shows that the PER was increased to ~12 dB with an improved modal purity of 90.1 %.

Next we chose to shape the input signal wavefront to obtain other selected vertically polarized scalar LP modes. Such modes can be represented as the in-phase and out-of-phase coherent superposition of two equally weighted vertically polarized OAM modes with opposite chirality ($LP_{a(b)} = |V, l\rangle \pm |V, -l\rangle$). The appropriate correlation filter was encoded to display the conjugate phase of the desired LP mode at the second SLM, and the iterative optimization procedure was again executed to obtain the target LP modes by controlling the input wavefront. We successfully generated the individual LP modes supported by the fibre as shown in **Fig. 2**(d). In addition, the orientation ($\alpha$) of the LP modes which can be interpreted in terms of a superposition of the two linearly polarized OAM modes with equal amplitude but different relative phase $\beta$ ($LP(\alpha) = |V, l\rangle + \exp(i\beta)|V, -l\rangle$, where $\alpha = \beta/2l$) can be completely controlled. One example of the measured intensity profiles of the rotated $LP_{11}$ output beams is shown in **Fig. 2**(e), indicating that the complex amplitudes of the four vector modes within the $LP_{11}$ group are well controlled. Note that all generated LP modes had an output power of ~10 W with a PER of >12 dB.

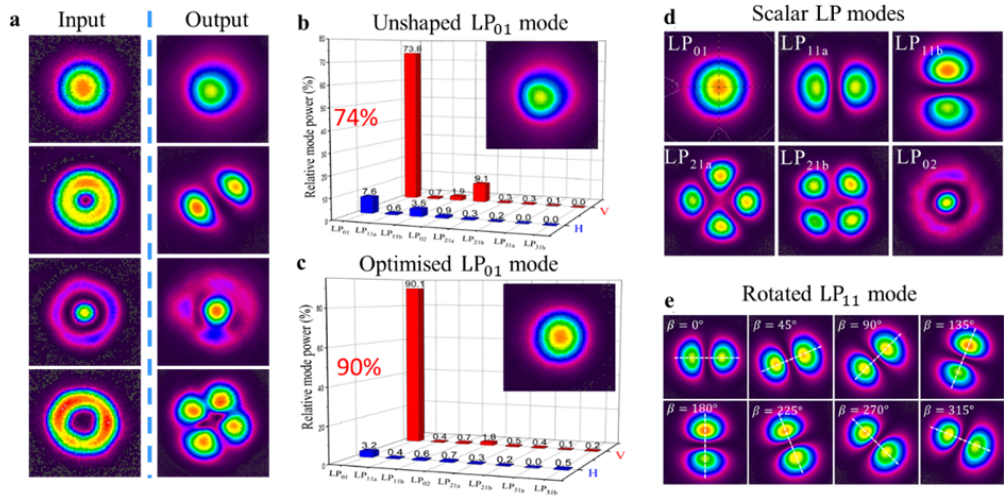

**Fig. 2 Scalar LP modes. a**, Examples of input and output beam profiles without input wavefront shaping; **b-c**, Modal decompositions of the unshaped $LP_{01}$ mode (b) and the optimised $LP_{01}$ mode (c); **d**, Measured intensity profiles of the different orders of output LP mode after input wavefront shaping; **e**, Measured output intensity profiles of the rotated $LP_{11}$ mode.

**Generation of vector fibre eigenmodes**

We further investigated the generation of degenerate fibre eigenmodes in the $LP_{11}$ and $LP_{21}$ groups supported by the fibre. Mathematically, such vector modes can be represented as an appropriate linear combination of the four OAM mode basis in the same LP group (the opposite chirality and orthogonal linear polarizations, i.e. $TM_{01} = (|H, l\rangle + |H, -l\rangle - i|V, l\rangle + i|V, -l\rangle)$, where $l = 1$). Thus, the conjugate phase of the desired vector mode can be encoded on the second SLM to construct an appropriate correlation filter. In order to get an unambiguous correlation signal, the PC in **Fig. 1** needs to be properly adjusted to ensure that the polarization of a horizontally polarized beam launched into the SMF is rotated to 45° with respect to the transmission axis of the fiberized PBS. This ensures that the correlation signal of the desired vector mode is solely transmitted to the horizontal polarization output port of the PBS and is measured by the power meter as the merit function during the iterative optimisation process. Note that there is a fixed slight path length difference (~100 μm) between the horizontal and vertical polarizations through the correlation filter system so that the relative phase between them should be calibrated before implementing the iterative optimisation process. Afterwards, the conjugate amplitude and phase of the desired vector mode was encoded in the second SLM, and by executing the adaptive wavefront shaping procedure, various fibre eigenmodes were successfully generated with high quality.

The second column in **Fig. 3**(a) shows the measured intensity distributions of the output vector beams including the radial polarization (the $TM_{01}$ mode), azimuthal polarization (the $TE_{01}$ mode), other cylindrically symmetric polarizations in the $LP_{11}$ group (the $HE_{21}^e$ and $HE_{21}^o$ modes) and $LP_{21}$ group (the $EH_{11}^e$ and $EH_{11}^o$ modes) at the maximum output power.

These are well matched to the theoretical intensity distributions as shown in the first column in **Fig. 3** (a). The cylindrically symmetric polarization states of the generated doughnut shaped modes were qualitatively confirmed by passing them through a rotatable linear polarizer. Rotated intensity profiles similar to the LP mode patterns were observed as predicted by theory and these are shown in the rest of the columns in **Fig. 3** (a). **Figure. 3** (b) shows the measured output power of the $TM_{01}$ and $EH_{11}^{e}$ modes as a function of pump power, which yielded a maximum average power of 12.0 W and 11.3 W, corresponding to slope efficiencies of 70 % and 66 %, respectively. This is comparable to the performance of conventional fibre amplifiers showing that a high power conversion efficiency can be achieved with our beam shaping approach. **Figure. 3** (c) shows the measured pulse shape of the generated vector modes with a pulse duration of ~150 ps full width at half maximum (FWHM), corresponding to an estimated peak power of ~12 kW. There is a smaller peak on the tail of the main pulse that was minimized by optimizing the diode temperature. The vector modal purity was analysed through a vector mode decomposition (see more details in the methods section). The results show that the fibre amplifier yielded a high vector mode purity of >91 % for all generated vector modes. **Figure. 3** (d) and (e) illustrate examples of the measured relative modal weightings for the generated $TM_{01}$ and $EH_{11}^e$ modes, respectively. We can see that the power of the undesired vector modes in the same LP mode group stayed below a level of −19 dB (1.3%) compared with the target vector mode. The slight fraction of power (<7 %) contained in the other order LP modes can be attributed to the residual mode coupling as well as the fast buildup of amplified stimulated emission (ASE) contained within the other spatial modes and which experienced a higher gain due to the undepleted population inversion under high pump power conditions.

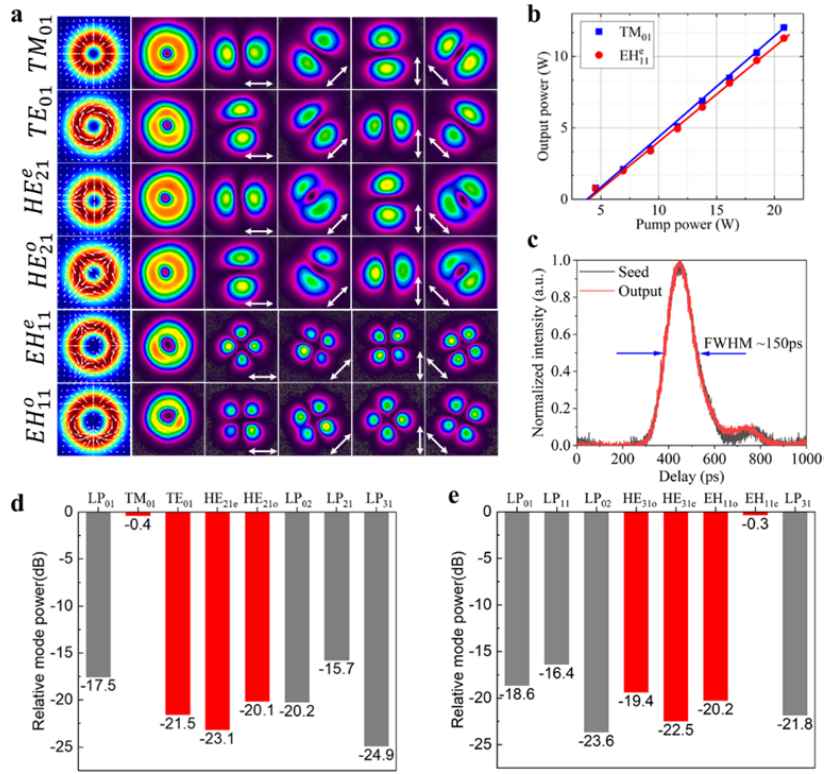

**Fig. 3 Cylindrical vector modes. a**, Theoretical (first column) and experimentally measured (second column) far-field intensity distributions of doughnut-shaped output vector modes and the intensity profiles of the corresponding mode passing through a linear polarizer at different orientations (white arrows) as a verification of the vector polarization state; **b**, Average output power of the $TM_{01}$ and $EH_{11e}$ modes versus the launched pump power; **c**, The measured temporal pulse shape having a pulse duration (FWHM) of ~150 ps; **d-e**, Examples of vector modal decomposition analysis for the $TM_{01}$ mode (d) and the $EH_{11e}$ mode (e), showing that >91 % of power is contained within the desired vector mode.

**Generation of linearly polarized OAM modes**

We finally investigated the controllable generation of linearly polarized OAM modes with tunable topological charges ($l = \pm 1, \pm 2$) at high output powers. The linearly polarized OAM modes can be represented as appropriate coherent combinations of two near-degenerate circularly polarized OAM modes with the same topological charge but opposite spin states (i.e. $|H, l\rangle = |\sigma^+, l\rangle + |\sigma^-, l\rangle$). The slight difference of propagation constant between the two components indicates that the combined linearly polarized OAM modes are actually not fibre eigenmodes so that the relative phase between them should be controlled to zero in order to get a purely linearly polarized OAM mode at the output. Since such linearly polarized OAM modes are the basis modes we employ to generate and analyse the spatial beams at the SLMs, the relevant correlation filter can be constructed just by encoding the conjugate phase mask of the desired basis mode on the second SLM. By implementing the iterative optimisation procedure we successfully generated various linearly polarized OAM modes with tunable topological charge and controllable polarization state. **Fig. 4**(a)-(d) shows four examples of output beam intensity profiles measured at an output power of ~10.5 W exhibiting pronounced doughnut shapes together with beam profiles of horizontal and vertical linear polarization components (shown in insets) indicating that the linear polarization states dominate the output beams. The helical phase fronts of the doughnut-shaped beams were confirmed by interfering them with a reference beam with a spherical wavefront, leading to the formation of spiral interference patterns. The number of spiral fringes and their rotation directions indicate that the target OAM modes with the desired topological charge were successfully achieved. The purity of the scalar OAM modes was analysed by decomposing the output beam into the OAM basis modes in both linear polarizations. **Fig. 4**(e) and (f) illustrate examples of measured relative modal weightings of two OAM modes $|V, 1\rangle$ and $|H, 2\rangle$ respectively, indicating that a high modal purity of >90 % was obtained. Note that switching between modes with different topological charge and polarization states was achieved by displaying the relevant phase mask on the second SLM and executing again the iterative optimisation procedures.

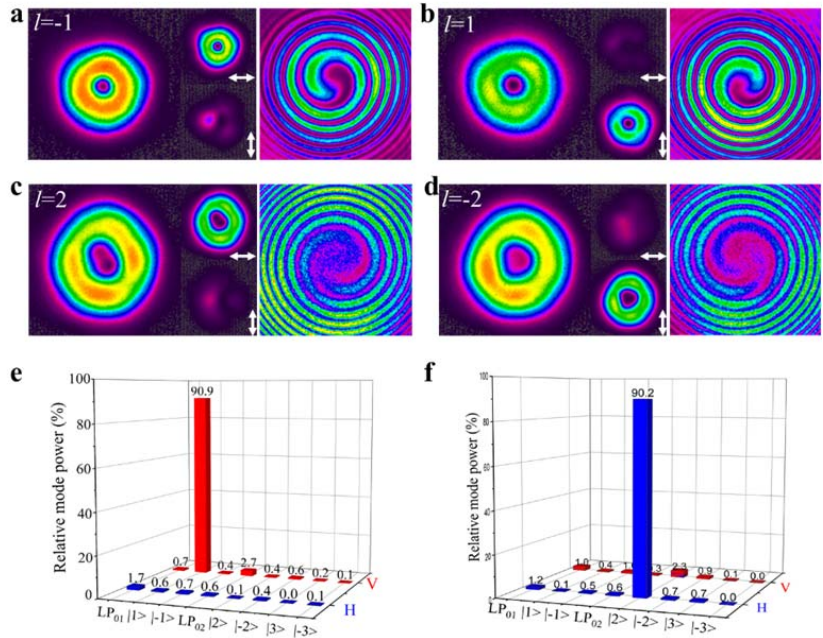

**Fig. 4 OAM modes. a-d**, The measured intensity profiles of the linearly polarized OAM output beams with topological charge of $l=-1$ (a), $l = 1$ (b), $l = 2$ (c) and $l = -2$ (d), respectively. The insets show the orthogonal components of the horizontal and vertical polarization states, indicating that the linear polarization dominates the output beams. The different spiral interference patterns indicate the output beams have the desired topological charges of the OAM state. **e-f**, Examples of modal decompositions results show that >90 % of the power is contained in the desired OAM mode of $|V, 1\rangle$ (e) and $|H, 2\rangle$ (f).

**Mode stability**

It is worth mentioning that the generated OAM modes and vector modes are quite stable and repeatable at fixed output power in the laboratory environment. The applied SLM phase pattern did not need to be changed once the optimisation process was completed for the target output beams and the beam quality was preserved for periods of at least several hours. **Figure. 5** (a) shows the measured modal purity of the $EH_{11}^e$ mode as a function of time at the maximum output power of ~11 W with an optimized SLM phase pattern. It shows that the average value of modal purity is 91.3 % with a ~6 % peak-to-valley fluctuation and a root mean square value of 0.012 over 60 minutes, confirming that a stable vector mode was generated. We noticed that each optimized input wavefront only works at a fixed output power, and that the beam quality decreases with variation in power around the value used for optimization. One example is shown in **Fig. 5** (b)-(d) which shows that the optimized doughnut-shaped $EH_{11}^e$ mode experiences severe beam distortion as the output power is decreased from 11 W to 2 W. However, the doughnut-shaped mode can be restored by running the algorithm again and reoptimizing the input wavefront as shown in **Fig. 5** (e). We attribute the mode evolution versus output power to changes in the mode transfer matrix due to changes in the thermal equilibrium of the fibre at different pump powers.

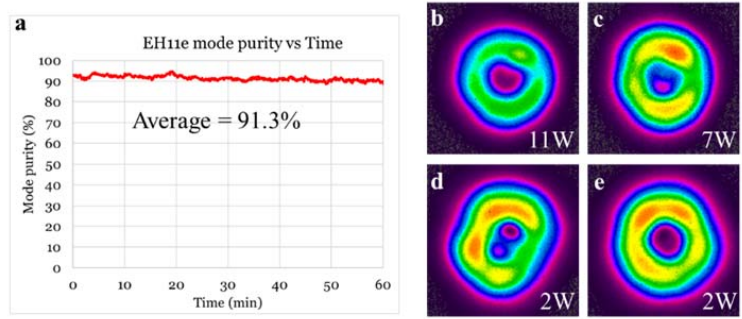

**Fig. 5 Modal stability measurement. a**, Measured modal purity of the $EH_{11}^e$ mode at a fixed output power of ~11 W as a function of time, indicating that a stable vector mode was generated and preserved over 60 minutes; **b-d**, mode evolution versus output power: the measured intensity profile of the optimised $EH_{11}^e$ mode at 11 W (b); the intensity profile of the output beam when the output power was decreased to 7 W (c) and 2 W (d) without changing the input wavefront; **e**, The intensity profile of the output beam at 2 W was restored to a doughnut shape after the input wavefront was re-shaped.

**Discussion and conclusion**

We have demonstrated a novel digital fibre amplifier capable of generating optical vortex beams of user specified topological charge and vector polarization states on demand at high power levels under electronic control. Our approach is based on adaptive shaping of the beam wavefront and polarization by means of a phase-only SLM placed at the input to a few-mode fibre amplifier. Using this approach, we successfully avoided optical damage to the SLM (by reducing the incident powers to below 250 mW) and were able to generate various spatial modes supported by the fibre including conventional scalar LP modes, scalar OAM modes and cylindrical vector modes. The amplifier produced 150 ps pulses at a repetition rate of 5.9 MHz with an average power >10 W corresponding to a peak power of above ~11 kW. The net gain of the final digital fibre amplifier stage reached ~>16dB. The output power might be further increased by a factor of ~3-5 if we increase the incident power to ~1.5 W at which point we reach the theoretical damage threshold of the SLM. As the SLM is expensive, and any contamination on the liquid display surface might reduce the damage threshold, we only operated the SLM at a power well below its damage threshold. With advances in SLM technology yielding a higher damage threshold and a faster refresh rate, the ability to operate at much higher output power and with faster optimisation times is to be expected. The demonstrated versatility in mode generation and power-handling capabilities of our approach opens a new pathway for the dynamic generation of high power structured light beams from simple and compact laser sources. This capability may be advantageous in a wide range of applications such as laser material processing, free-space optical communication and high-resolution imaging.

**Acknowledgements**

This work is funded by the UK EPSRC through project EP/P012248/1; EP/P027644/1; and EP/P030181/1. J.C. is supported from the Discovery (DP170101400, DE180100009) program of the Australian Research Council (ARC).

**Author contributions**

D.L. designed and performed the experiments, and analysed the data. D.L. and J.C. devised the arbitrary beam generator and the digital correlation filter. Y.F. did simulations on transverse modes amplifications in the fibre amplifier. Y.J., S.A. and D.J.R. participated in scientific discussions. D.J.R. proposed and supervised the work. All authors contributed to paper preparation.

**Competing interests**

The authors declare no competing interests.

**Methods**

**Experimental setup.** A gain-switched laser diode operating at a wavelength of 1035 nm and emitting ~150 ps Gaussian-shaped pulses at a repetition rate of 5.9 MHz is used as the seed to a fibre based Master Oscillator Power Amplifier (MOPA) system. The pulses are first pre-amplified to an average power of ~250 mW in a chain of PM-SMF amplifiers. The output fibre of the final pre-amplifier stage has a core diameter of 10 μm with a numerical aperture (NA) of 0.085, ensuring delivery of a pure fundamental mode with a nearly Gaussian-shaped intensity profile and a well-defined flat phase profile. The resulting amplified beam is then collimated with a spot size of ~6.8 mm and injected onto a reflective phase-only liquid crystal SLM (Holoeye PLUTO-2-NIR-149, AG, Germany) through some polarization diversity optics. The SLM has an active area of 15.36×8.64 mm comprising 1920×1080 pixels at a pitch of 8 μm. It can operate at any wavelength from 1000 nm to 1100 nm with a reflectivity of 93 %. The area of the SLM is divided into two halves to display the phase masks and this enables us to generate any spatial and polarization state (in amplitude and phase) in the Fourier plane. Although illustrated in a transmissive topology for clarity in **Fig. 1,** in practice each incident beam reflects off its own phase mask displayed on the SLM.

The re-shaped beam is first coupled to a ~10 m length of passive single-clad fibre with a refractive index profile matched to that of the commercially available step-index FM-LMA Yb-doped fibre (Liekki Yb1200-30/250DC) used in the final stage amplifier. The passive single-clad fiber is used as an aperture that effectively blocks any forward propagated unwanted light that otherwise might be amplified, and also work with the free-space isolator to block any backward propagated amplified spontaneous emissions (ASE) to protect the SLM. The resulting signal is passed through a home-made free-space polarization independent isolator which ensures a reduced polarization dependent loss and is injected into the final stage amplifier fibre which has a core diameter of 30 μm with an NA of ~0.075 ± 0.005 and a cladding diameter of 250 μm with an NA of 0.46. The fibre has a peak nominal cladding absorption of ~14 dB/m at 976 nm, and ~1.5 m length of fibre was used in the experiment. Both ends of the fibre were spliced to ~1 mm long silica end-caps with a diameter of 250 μm to suppress any potential parasitic lasing. The input end facet was perpendicularly cleaved but the output end facet was angle polished with an angle of ~8 degrees to further reduce the back-reflection and unwanted optical feedback.

At the fibre amplifier output, a fraction of the output beam is characterized by a digital spatial mode correlation filter based on a second SLM, and the resulting correlation signal is used in an iterative optimization process to adjust the wavefront and polarization of the beam launched into the fibre as needed to generate the target beam at the output. Note that we used a basis set of linearly polarized modes with helical wavefronts (and with intensity profiles matched to those of the fibre) to generate and analyse the spatial beam profile at the SLMs. Any scalar or vector fibre mode can thus be described in terms of a superposition of modes in this basis.

**Vector modal purity measurement.** To analyse the modal purity of the generated vector modes, a vector modal decomposition technique is implemented. First, the output beam is decomposed into the orthogonal linearly polarized OAM mode basis through an inner product measurement as follows:

$$c_{\varepsilon,L} = \langle \Psi_{\varepsilon,L} | E_{out} \rangle \tag{2}$$

where $\Psi_{\varepsilon,L}$ represents the horizontal OAM basis mode ($|H, \pm L\rangle$) for $\varepsilon = 1$, and vertical polarization OAM basis mode ($|V, \pm L\rangle$) for $\varepsilon = -1$, respectively ($L \leq 3$ in practice); The $c_{\varepsilon,L}^2$ are the corresponding modal weights which are proportional to the correlation signal of the on-axis intensity coupled to the SMF. The relative modal weight of each LP group can be derived as the sum of the modal weights of the OAM mode basis with the same order of L and orthogonal polarizations as follows:

$$\rho_L = \frac{\sum c_{\varepsilon,\pm L}^2}{\sum_{l=0}^{l=3} c_{\varepsilon,\pm |l|}^2} \tag{3}$$

The next step is to measure the relative modal weight of each degenerate vector mode within the same LP group, which is also achieved by an inner product measurement. Consider the case where the output is the pure $HE_{L+1}^{even}$ mode, then the inner product with the conjugate phase of each degenerate vector mode has the following form:

$$\begin{cases} \langle HE_{L+1,m}^{even} | HE_{L+1,m}^{even} \rangle = 2 \begin{pmatrix} 1 \\ 1 \end{pmatrix} |0\rangle + \begin{pmatrix} 1 \\ -1 \end{pmatrix} |2L\rangle + \begin{pmatrix} 1 \\ -1 \end{pmatrix} |-2L\rangle \\ \langle HE_{L+1,m}^{odd} | HE_{L+1,m}^{even} \rangle = i \begin{pmatrix} -1 \\ 1 \end{pmatrix} |2L\rangle + i \begin{pmatrix} 1 \\ -1 \end{pmatrix} |-2L\rangle \\ \langle EH_{L-1,m}^{even} | HE_{L+1,m}^{even} \rangle = 2 \begin{pmatrix} 1 \\ -1 \end{pmatrix} |0\rangle + \begin{pmatrix} 1 \\ 1 \end{pmatrix} |2L\rangle + \begin{pmatrix} 1 \\ 1 \end{pmatrix} |-2L\rangle \\ \langle EH_{L-1,m}^{odd} | HE_{L+1,m}^{even} \rangle = -i \begin{pmatrix} 1 \\ 1 \end{pmatrix} |2L\rangle + i \begin{pmatrix} 1 \\ 1 \end{pmatrix} |-2L\rangle \end{cases} \tag{4}$$

The SMF located on the optical axis in the far-field can only detect the non-OAM component ($|0\rangle$) of the electric field within the resulting correlation signal. It can be seen that the inner product of the conjugate phase of the $HE_{L+1}^{even}$ mode contains an on-axis intensity component with a linear polarization aligned to 45° with respect to the horizontal direction. The inner product of the conjugate phase of the $EH_{L-1}^{even}$ mode contains the on-axis intensity component as well, but with orthogonal linear polarization (-45°). To differentiate between these two signals, a PC is adjusted to rotate the linear polarization of the correlation signal of the $HE_{L+1}^{even}$ phase mask to the horizontal direction, which is aligned to the 1st output port of the PBS. Correspondingly, the linear polarization of the correlation signal of the $\text{EH}_{L-1}^{\text{even}}$ phase mask is rotated to the vertical direction, which is aligned to the 2$^\text{nd}$ output port of the PBS. By sequentially displaying the conjugate phase of each vector mode, the measured power ($P_j$) of the 1$^\text{st}$ output port of the PBS can be considered as the weight of each vector mode. Then the relative weight ($\eta_j$) of each vector mode can be expressed as:

$$\eta_j = \frac{P_j}{\sum P_j} \tag{5}$$

where $j$ represents the $j$-th vector mode in the LP group (each LP group contains four degenerate vector modes). As a result, the overall relative weight of each vector mode ($\gamma_{Lj}$) can be expressed as:

$$\gamma_{Lj} = \eta_j \rho_L \tag{6}$$

**Reference**

1. D. G. Grier, "A revolution in optical manipulation," Nature **424**, 810-816 (2003).
2. M. Padgett and R. Bowman, "Tweezers with a twist," Nat. Photon. **5**, 343-348 (2011).
3. J. Wang, J. Y. Yang, I. M. Fazal, N. Ahmed, Y. Yan, H. Huang, Y. X. Ren, Y. Yue, S. Dolinar, M. Tur, and A. E. Willner, "Terabit free-space data transmission employing orbital angular momentum multiplexing," Nat. Photon. **6**, 488-496 (2012).
4. N. Bozinovic, Y. Yue, Y. X. Ren, M. Tur, P. Kristensen, H. Huang, A. E. Willner, and S. Ramachandran, "Terabit-scale orbital angular momentum mode division multiplexing in fibers," Science **340**, 1545-1548 (2013).
5. J. Hamazaki, R. Morita, K. Chujo, Y. Kobayashi, S. Tanda, and T. Omatsu, "Optical-vortex laser ablation," Opt. Express **18**, 2144-2151 (2010).
6. R. Weber, A. Michalowski, M. Abdou-Ahmed, V. Onuseit, V. Rominger, M. Kraus, and T. Graf, "Effects of Radial and Tangential Polarization in Laser Material Processing," Phys. Procedia **12**, 21-30 (2011).
7. L. J. Wong, "Direct acceleration of an electron in infinite vacuum by a pulsed radially-polarized laser beam," Opt. Express **18**, 25035-22501 (2010).
8. X. A. Hao, C. F. Kuang, T. T. Wang, and X. Liu, "Effects of polarization on the de-excitation dark focal spot in STED microscopy," J. Opt. **12**, 115707 (2010).
9. S. Furhapter, A. Jesacher, S. Bernet, and M. Ritsch-Marte, "Spiral phase contrast imaging in microscopy," Opt. Express **13**, 689-694 (2005).
10. M. W. Beijersbergen, R. P. C. Coerwinkel, M. Kristensen, and J. P. Woerdman, "Helical-wavefront laser beams produced with a spiral phaseplate," Opt. Commun. **112**, 321-327 (1994).
11. N. R. Heckenberg, R. McDuff, C. P. Smith, and A. G. White, "Generation of optical phase singularities by computer-generated holograms," Opt. Lett. **17**, 221-223 (1992).
12. G. Gibson, J. Courtial, M. Padgett, M. Vasnetsov, V. Pas'ko, S. Barnett, and S. Franke-Arnold, "Free-space information transfer using light beams carrying orbital angular momentum," Opt. Express **12**, 5448-5456 (2004).
13. G. Milione, H. I. Sztul, D. A. Nolan, and R. R. Alfano, "Higher-Order Poincare Sphere, Stokes Parameters, and the Angular Momentum of Light," Phys. Rev. Lett. **107**, 053601 (2011).
14. L. Marrucci, C. Manzo, and D. Paparo, "Optical spin-to-orbital angular momentum conversion in inhomogeneous anisotropic media," Phys. Rev. Lett. **96**, 163905 (2006).
15. M. Beresna, M. Gecevicius, P. G. Kazansky, and T. Gertus, "Radially polarized optical vortex converter created by femtosecond laser nanostructuring of glass," Appl. Phys. Lett. **98**, 201101 (2011).
16. N. F. Yu, P. Genevet, M. A. Kats, F. Aieta, J. P. Tetienne, F. Capasso, and Z. Gaburro, "Light Propagation with Phase Discontinuities: Generalized Laws of Reflection and Refraction," Science **334**, 333-337 (2011).
17. D. Lin, J. M. O. Daniel, and W. A. Clarkson, "Controlling the handedness of directly excited Laguerre-Gaussian modes in a solid-state laser," Opt. Lett. **39**, 3903-3906 (2014).
18. A. Loescher, J. P. Negel, T. Graf, and M. A. Ahmed, "Radially polarized emission with 635 W of average power and 2.1 mJ of pulse energy generated by an ultrafast thin-disk multipass amplifier," Opt. Lett. **40**, 5758-5761 (2015).
19. H. Sroor, I. Litvin, D. Naidoo, and A. Forbes, "Amplification of higher order Poincare sphere beams through Nd:YLF and Nd:YAG crystals," Appl. Phys. B **125**, 49 (2019).
20. D. Lin, J. M. O. Daniel, M. Gecevicius, M. Beresna, P. G. Kazansky, and W. A. Clarkson, "Cladding-pumped ytterbium-doped fiber laser with radially polarized output," Opt. Lett. **39**, 5359-5361 (2014).
21. D. Lin, N. Baktash, S. U. Alam, and D. J. Richardson, "106 W, picosecond Yb-doped fiber MOPA system with a radially polarized output beam," Opt. Lett. **43**, 4957-4960 (2018).
22. D. Stellinga, M. E. Pietrzyk, J. M. E. Glackin, Y. Wang, A. K. Bansal, G. A. Turnbull, K. Dholakia, I. D. W. Samuel, and T. F. Krauss, "An Organic Vortex Laser," Acs Nano **12**, 2389-2394 (2018).
23. M. A. Ahmed, J. Schulz, A. Voss, O. Parriaux, J. C. Pommier, and T. Graf, "Radially polarized 3 kW beam from a $CO_2$ laser with an intracavity resonant grating mirror," Opt. Lett. **32**, 1824-1826 (2007).
24. S. Ngcobo, I. Litvin, L. Burger, and A. Forbes, "A digital laser for on-demand laser modes," Nat. Commun. **4**, 2289 (2013).
25. D. Naidoo, F. S. Roux, A. Dudley, I. Litvin, B. Piccirillo, L. Marrucci, and A. Forbes, "Controlled generation of higher-order Poincare sphere beams from a laser," Nat. Photon. **10**, 327-333 (2016).
26. H. Sroor, Y. W. Huang, B. Sephton, D. Naidoo, A. Valles, V. Ginis, C. W. Qiu, A. Ambrosio, F. Capasso, and A. Forbes, "High-purity orbital angular momentum states from a visible metasurface laser," Nat. Photon. (2020).
27. N. C. Zambon, P. St-Jean, M. Milicevic, A. Lemaitre, A. Harouri, L. Le Gratiet, O. Bleu, D. D. Solnyshkov, G. Malpuech, I. Sagnes, S. Ravets, A. Amo, and J. Bloch, "Optically controlling the emission chirality of microlasers," Nat. Photon. **13**, 283-289 (2019).

28. M. J. Strain, X. L. Cai, J. W. Wang, J. B. Zhu, D. B. Phillips, L. F. Chen, M. Lopez-Garcia, J. L. O'Brien, M. G. Thompson, M. Sorel, and S. Y. Yu, "Fast electrical switching of orbital angular momentum modes using ultra-compact integrated vortex emitters," Nat. Commun. **5**, 4856 (2014).
29. X. L. Cai, J. W. Wang, M. J. Strain, B. Johnson-Morris, J. B. Zhu, M. Sorel, J. L. O'Brien, M. G. Thompson, and S. T. Yu, "Integrated Compact Optical Vortex Beam Emitters," Science **338**, 363-366 (2012).
30. P. Miao, Z. F. Zhang, J. B. Sun, W. Walasik, S. Longhi, N. M. Litchinitser, and L. Feng, "Orbital angular momentum microlaser," Science **353**, 464-467 (2016).
31. D. Lin, J. Carpenter, Y. T. Feng, S. Jain, Y. M. Jung, Y. J. Feng, M. N. Zervas, and D. J. Richardson, "Reconfigurable structured light generation in a multicore fibre amplifier," Nat. Commun. **11**, 3986 (2020).
32. B. Ndagano, R. Brüning, M. McLaren, M. Duparré, and A. Forbes, "Fiber propagation of vector modes," Opt. Express **23**, 17330-17336 (2015).
33. S. Ramachandran, P. Kristensen, and M. F. Yan, "Generation and propagation of radially polarized beams in optical fibers," Opt. Lett. **34**, 2525-2527 (2009).
34. R. S. Chen, J. H. Wang, X. Q. Zhang, A. T. Wang, H. Ming, F. Li, D. Chung, and Q. W. Zhan, "High efficiency all-fiber cylindrical vector beam laser using a long-period fiber grating," Opt. Lett. **43**, 755-758 (2018).
35. T. Wang, F. Shi, Y. P. Huang, J. X. Wen, Z. Q. Luo, F. F. Pang, T. Y. Wang, and X. L. Zeng, "High-order mode direct oscillation of few-mode fiber laser for high-quality cylindrical vector beams," Opt. Express **26**, 11850-11858 (2018).
36. Y. M. Jung, Q. Y. Kang, R. Sidharthan, D. Ho, S. Yoo, P. Gregg, S. Ramachandran, S. U. Alam, and D. J. Richardson, "Optical Orbital Angular Momentum Amplifier Based on an Air-Hole Erbium-Doped Fiber," J. Light. Technol. **35**, 430-436 (2017).
37. J. W. Ma, F. Xia, S. Chen, S. H. Li, and J. Wang, "Amplification of 18 OAM modes in a ring-core erbium-doped fiber with low differential modal gain," Opt. Express **27**, 38087-38097 (2019).
38. I. N. Papadopoulos, S. Farahi, C. Moser, and D. Psaltis, "High-resolution, lensless endoscope based on digital scanning through a multimode optical fiber," Biomed. Opt. Express **4**, 260-270 (2013).
39. S. M. Popoff, G. Lerosey, R. Carminati, M. Fink, A. C. Boccara, and S. Gigan, "Measuring the Transmission Matrix in Optics: An Approach to the Study and Control of Light Propagation in Disordered Media," Phys. Rev. Lett. **104**(2010).
40. R. Di Leonardo and S. Bianchi, "Hologram transmission through multi-mode optical fibers," Opt. Express **19**, 247-254 (2011).
41. T. Cizmar and K. Dholakia, "Shaping the light transmission through a multimode optical fibre: complex transformation analysis and applications in biophotonics," Opt. Express **19**, 18871-18884 (2011).
42. M. Ploschner, T. Tyc, and T. Cizmar, "Seeing through chaos in multimode fibres," Nat. Photon. **9**, 529-+ (2015).
43. J. Carpenter, B. J. Eggleton, and J. Schroder, "Observation of Eisenbud-Wigner-Smith states as principal modes in multimode fibre," Nat. Photon. **9**, 751-758 (2015).
44. G. Lopez-Galmiche, Z. S. Eznaveh, J. E. Antonio-Lopez, A. M. V. Benitez, J. R. Asomoza, J. J. S. Mondragon, C. Gonnet, P. Sillard, G. Li, A. Schulzgen, C. M. Okonkwo, and R. A. Correa, "Few-mode erbium-doped fiber amplifier with photonic lantern for pump spatial mode control," Opt. Lett. **41**, 2588-2591 (2016).
45. J. Montoya, C. Aleshire, C. Hwang, N. K. Fontaine, A. Velazquez-Benitez, D. H. Martz, T. Y. Fan, and D. Ripin, "Photonic lantern adaptive spatial mode control in LMA fiber amplifiers," Opt. Express **24**, 3405-3413 (2016).
46. R. Florentin, V. Kermene, J. Benoist, A. Desfarges-Berthelemot, D. Pagnoux, A. Barthelemy, and J. P. Huignard, "Shaping the light amplified in a multimode fiber," Light Sci. Appl. **6**, e16208 (2017).
47. M. Pasienski and B. DeMarco, "A high-accuracy algorithm for designing arbitrary holographic atom traps," Opt. Express **16**, 2176-2190 (2008).
48. S. H. Tao and W. X. Yu, "Beam shaping of complex amplitude with separate constraints on the output beam," Opt. Express **23**, 1052-1062 (2015).
49. D. Flamm, D. Naidoo, C. Schulze, A. Forbes, and M. Duparre, "Mode analysis with a spatial light modulator as a correlation filter," Opt. Lett. **37**, 2478-2480 (2012).
50. J. Carpenter, B. C. Thomsen, and T. D. Wilkinson, "Degenerate Mode-Group Division Multiplexing," J. Light. Technol. **30**, 3946-3952 (2012).